\documentclass[11pt,a4paper]{article}
\usepackage{jheppub_kim}
\usepackage{pdflscape}
\usepackage{amsmath}
\usepackage{amssymb}
\usepackage{dcolumn}
\usepackage{bm}
\usepackage{color}
\usepackage{epsfig}
\usepackage{amsfonts}
\usepackage{graphicx}
\usepackage{subfigure}
\usepackage{dcolumn}
\begin{document}

\title{Gravitational Baryogenesis in Ho$\check{r}$ava-Lifshitz gravity}

\author[a]{Sayani Maity,}

\author[b]{~Prabir Rudra {\footnote{Corresponding Author}}}

\affiliation[a] {Department of Mathematics, Techno India Salt
Lake, Sector-V, Kolkata-700 091, India.}

\affiliation[b]{Department of Mathematics, Asutosh College,
Kolkata-700 026, India.}{}

\emailAdd{sayani.maity88@gmail.com}

\emailAdd{~prudra.math@gmail.com}

\abstract{In this work we intend to address the matter-antimatter
asymmetry via the gravitational baryogenesis mechanism in the
background of a quantum theory of gravity. We investigate this
mechanism under the framework of Ho$\check{r}$ava-Lifshitz
gravity. We will compute the baryon-to-entropy ratio in the chosen
framework and investigate its physical viability against the
observational bounds. We also conduct the above study for various
sources of matter like scalar field and Chaplygin gas as specific
examples. We speculate that quantum corrections from the
background geometry will lead to interesting results.}


\maketitle
\section{\normalsize\bf{Introduction}}
Cosmic Microwave Background observations \cite{benett1} and Big
Bang nucleosynthesis  predictions \cite{burles1} have confirmed
the presence of excess matter over antimatter in the universe.
There has been a lot of debate over such observations and
gravitational baryogenesis have been proposed as the most suitable
mechanism for such asymmetry \cite{1DHKRKGDMHSPJ04, Lambiase1,
Lambiase2, Li1}. Baryogenesis mechanism incorporates one of
Sakharov's criteria \cite{Sakharov1}. According to him in the process of generation of matter-antimatter asymmetry the following three conditions must be satisfied: baryon number non-conservation, C and CP symmetry violation and deviation from thermal equilibrium.  \\

The baryon asymmetry gravitational baryogenesis term used in
\cite{1DHKRKGDMHSPJ04} is of the form
\begin{equation}
\frac{1}{M_{*}^{2}}\int d^{4}x \sqrt{-g}(\partial_{\mu}R) J^{\mu},
\end{equation}
where the parameter $M_{*}$ is the cut-off scale of the underlying
effective gravitational theory. $J^{\mu}$, $g$ and
$R=12H^{2}+6\dot{H}$ stand for the baryonic matter current, the
trace of the metric tensor and the Ricci scalar respectively. This
is a CP-violating interaction term which can be acquired from
higher order interactions in the fundamental gravitational theory
\cite{1DHKRKGDMHSPJ04}. If flat FRW geometry is considered then
baryon-to-entropy ratio $\eta_{B}/s$ is proportional to $\dot{R}$.
Further the baryon-to-entropy ratio becomes zero when the matter
fluid corresponds to relativistic matter with EOS parameter
$\omega=1/3$. The predicted baryon-to-entropy ratio is
$\eta_{B}/s\simeq 9.2 \times 10^{-11}$ \cite{1BCL03}.\\

With the discovery of the accelerated expansion of the universe
\cite{Perlmutter, Spergel}, Einstein's theory of general
relativity needed serious revisions and hence different dynamical dark energy models and modified theories of
gravity came to the foreground. Moreover there has been a
prolonged attempt to reconcile the physics of the large with the
physics of the small which will result in a quantum theory of
gravity. Ho$\check{r}$ava-Lifshitz gravity is a novel attempt
towards such a theory of quantum gravity \cite{Horava1, Horava2,
Horava3}. The theory is devoid of full diffeomorphism invariance
but it has UV completeness. Not only that, the theory has a three
dimensional general covariance and time re-parameterization
invariance. In fact it is a non-relativistic renormalizable
quantum theory of gravity possessing higher order spatial
derivatives \cite{Horava4, Horava5}. The singularity problem has
plagued most of the basic theories of cosmology over the years.
The singularity at the beginning of the universe and the
singularity inside a black hole has remained a totally unknown and
unexplained fact. One of the basic motivations of the
Ho$\check{r}$ava-Lifshitz gravity is to provide an explanation to
the singularity paradigm. Recently there have been extensive
research in Ho$\check{r}$ava-Lifshitz gravity \cite{Khodadi1,
Hartong, Garattini, Prabir1, Prabir2}.\\

Various dark energy candidates have been proposed to explain the
late time acceleration. The cosmological constant is the most
common candidate that plays the role of dark energy. Various other
models of dark energy have been proposed over the years without
using the cosmological constant. There is one class of dynamical
dark energy models which includes a scalar field \cite{lucchin,
scalar}. The introduction of the scalar field $\phi$ makes the
vacuum energy dynamical and the model can represent a wide range
of cosmological scenarios from inflation to late time
acceleration. In the scalar field models $\phi$ is assumed to be
spatially homogeneous, $\dot{\phi}^2/2$ is the kinetic energy and
$V(\phi)$ is the potential energy \cite{scalar}.  Scalar field
models of DE have been discussed in \cite{scalar2, scalar3,
scalar4, scalar5}. Chaplygin gas model \cite{chaplygin1,
chaplygin2} is another form of dark energy proposed in literature.
From the initially developed pure Chaplygin gas it gradually
evolved to generalized Chaplygin gas (GCG). The GCG model has been
discussed extensively in literature \cite{chap1, chap2, chap3,
chap4, chap5, chap6} as an effective model of dark energy for
interpreting the accelerating universe. Interestingly, it
describes two unknown dark sections of the universe- dark energy
and dark matter in a single matter component \cite{1BNTGBVRD01}.
It is characterized by an exotic equation of state
$p_{GCG}=-\frac{A}{\rho_{GCG}^{\alpha}}$, where $0<\alpha\leq 1$
and A is a positive constant \cite{1BMCBOSAA02}. For $\alpha=1$,
it gives original Chaplygin gas and $\alpha=0$ accounts for
$\Lambda$CDM model. To constrain the parameter space, GCG model
has been investigated with different cosmic observational data set
such as the baryon acoustic oscillation (BAO), microwave
background radiation (CMBR), geometric
information from SN Ia \cite{1XLLJ10,1LJGYXL09,1LZWPYH09,1WPYH07}.\\

Gravitational baryogenesis has been studied under the framework of
different gravity theories over the past few years. In 2006
baryogenesis was investigated in the framework of $f(R)$ gravities
in \cite{Lambiase1}. In \cite{1PMPLPPJ17} Ramos and Paramos
generalized this model by introducing a non-minimal coupling (NMC)
between curvature and matter and investigated the impact of NMC on
the mechanism of gravitational baryogenesis. Aghamohammadi and
Hossienkhani in \cite{1AAHH17} considered an anisotropic metric
and investigated its effect on the baryon to entropy ratio in the
context of $f(R)$ gravity. Odintsov et al in \cite{Odintsov1}
studied gravitational baryogenesis in Loop quantum cosmology. This
is the first instance where the mechanism of baryogenesis was
investigated under the framework of a theory of quantized gravity.
Baryogenesis was investigated in $f(T)$ gravity by Oikonomou et al
in \cite{Oikono1}. Gauss-Bonnett gravitational baryogenesis is
studied in \cite{1OSDOVK16}. Recently in \cite{1HZCQ17}, Huang and Cai
introduced gravitational baryogenesis mechanism in the vacuum inflation
model and showed that this model can produce acceptable baryon asymmetry.
In \cite{1SJSAR17}, Sakstein and Solomon pointed out the importance of
Lorentz-violating gravity theories that may yield matter-antimatter
asymmetry consistent with the observational bound.\\

Drawing motivation from \cite{Odintsov1} we would like to
investigate the gravitational baryogenesis mechanism in the
framework of Ho$\check{r}$ava-Lifshitz gravity in this note.
Section 2, is devoted to study the salient features of
Ho$\check{r}$ava-Lifshitz gravity and baryogenesis in its
framework. In this section, we also quantify the results and study
the the values of the parameters for which the results can be
compatible with the observational bounds. Section 3 is focussed to
discuss the effect of scalar field on the gravitational
baryogenesis in the framework of Ho$\check{r}$ava-Lifshitz
gravity. In section 4 we will study baryogenesis with dark energy
in the form generalized Chaplygin gas. Finally the paper ends with
some concluding remarks in section 5.

\section{\normalsize\bf{Baryogenesis mechanism in the background of Ho$\check{r}$ava-Lifshitz Gravity}}

Here we consider a Friedmann-Robertson-Walker (FRW) background
geometry with metric
\begin{equation}
\mbox{d}s^{2}=
c^{2}\mbox{d}t^{2}-\mbox{d}\sigma^{2}=c^{2}\mbox{d}t^{2}-a^{2}\Big{[}\frac{\mbox{d}r^{2}}{1-kr^{2}}+r^{2}\Big{(}\mbox{d}\theta^{2}+\sin^{2}\theta
\mbox{d}\phi^{2}\Big{)}\Big{]}
\end{equation}
where $a(t)$ is known as scale factor or expansion factor. Here the curvature, $k=-1,0,1$ represents open, flat and closed universe respectively.\\

Einstein's field equations for Ho$\check{r}$ava-Lifshitz gravity
\cite{1JMSEN10, 1BSDU11} are given by
\begin{equation}
H^{2}+\frac{k}{a^{2}}=\frac{8\pi
G}{3}\rho+\frac{k^{2}}{2\Lambda
a^{4}}+\frac{\Lambda}{2}
\end{equation}
and
\begin{equation}
\dot{H}+\frac{3}{2}H^{2}+\frac{k}{2a^{2}}=-4\pi G
p-\frac{k^{2}}{4\Lambda a^{4}}+\frac{3\Lambda}{4}
\end{equation}
where $\rho$ and $p$ are energy density and pressure of the
universe, $H=\frac{\dot{a}}{a}$ is the Hubble parameter (choosing
$c=1$), $\Lambda$ is the cosmological
constant and $G$ is the cosmological Newton's constant.\\
The energy density satisfies the continuity equation
\begin{equation}
\dot{\rho}+3H(\rho+p)=0
\end{equation}
From the above equation we get
\begin{equation}
\rho=\rho_{0}a^{-3(1+\omega)}
\end{equation}
where $\rho_{0}$ is the integration constant and $\omega=p/\rho$
is the EoS parameter.

The baryon-to-entropy ratio as given in \cite{1DHKRKGDMHSPJ04} is
\begin{equation}
\frac{\eta_{B}}{s}\simeq-\frac{15g_{b}}{4\pi^{2}g_{*}}\frac{\dot{R}}{M_{*}^{2}T}\mid_{T_{D}}
\end{equation}
where $g_{b}$ is the number of intrinsic degrees of freedom of the baryons,
$T_{D}$ is the critical temperature of the universe at which the baryon asymmetry generating interactions occur.
$g_{*}$ is the total degrees of freedom of effective massless particles ($g_{*}\sim106$).\\
We shall assume that a thermal equilibrium exists and the universe
evolves slowly from an equilibrium state to another equilibrium
state. In this process the energy density is related to the
temperature of each state by
\begin{equation}
\rho=\frac{\pi^{2}}{30}g_{*}T^{4}
\end{equation}
In the standard Einstein-Hilbert gravity framework, if the
universe is filled with a perfect matter fluid with constant
equation of state $\omega=p/\rho$, the Ricci scalar takes the form
\cite{1OSDOVK16}
\begin{equation}
R=-8\pi G(1-3\omega)\rho
\end{equation}
Using equations (2.2) and (2.3) the Ricci scalar reads
\begin{equation}
R=8\pi G(1-3\omega)\rho+\frac{21\Lambda}{2}-\frac{6k}{a^{2}}
\end{equation}
\begin{equation}
\dot{R}=8\pi G(1-3\omega)\dot{\rho}+\frac{12k}{a^{3}}\dot{a}
\end{equation}
Using the above result in equation (2.6), we obtain the
baryon-to-entropy ratio for Ho$\check{r}$ava-Lifshitz gravity.

For a flat universe, Einstein's field equation (2.2) for
Ho$\check{r}$ava-Lifshitz gravity can be analytically solved by
using equation (2.5) to yield the scale factor
\begin{equation}
a(t)=\frac{16\pi}{3}^\frac{1}{3(1+\omega)}\left(\left(\frac{G
\rho_{0}}{\Lambda}\right)^\frac{1}{6}\mbox{sinh}\left[\frac{3}{4}(1+\omega)\sqrt{\Lambda}(\sqrt{2}t+2\sqrt{3}C_{1})\right]^\frac{1}{3}\right)^\frac{2}{1+\omega}
\end{equation}
where $C_{1}$ is the integration constant.
Correspondingly the Hubble's parameter takes the form
\begin{equation}
H(t)=\sqrt{\frac{\Lambda}{2}}
\mbox{coth}\left[\frac{3}{4}(1+\omega)\sqrt{\Lambda}(\sqrt{2}t+2\sqrt{3}C_{1})\right]
\end{equation}
The energy density as a function of cosmic time is obtained as
\begin{equation}
\rho(t)=\frac{3\Lambda}{16\pi G}\left(
\mbox{sinh}\left[\frac{3}{4}(1+\omega)\sqrt{\Lambda}(\sqrt{2}t+2\sqrt{3}C_{1})\right]^\frac{1}{3}\right)^{-6}
\end{equation}

Using equations (2.7) and (2.13) we obtain the decoupling time
$t_{D}$ as a function of critical temperature $T_{D}$ as follows
\begin{equation}
t_{D}=\frac{2\sqrt{2}}{3(1+\omega)\sqrt{\Lambda}}\mbox{arcsinh}\left[\frac{3}{2\pi}\sqrt{\frac{5\Lambda}{2\pi
g_{\ast}G T_{D}^{4}}}\right]
\end{equation}
By using equation (2.13) in equation (2.10) for flat universe, we
get $\dot{R}$ in terms of cosmic time as follows
\begin{eqnarray*}
\dot{R}=2^{-\frac{3}{2}}9(1+\omega)(3\omega-1)\Lambda^{\frac{3}{2}}\mbox{coth}\left[\frac{3}{4}(1+\omega)\sqrt{\Lambda}(\sqrt{2}t+2\sqrt{3}C_{1})\right]\times
\end{eqnarray*}
\begin{equation}
\left(
\mbox{sinh}\left[\frac{3}{4}(1+\omega)\sqrt{\Lambda}(\sqrt{2}t+2\sqrt{3}C_{1})\right]^\frac{1}{3}\right)^{-6}
\end{equation}

Using equation (2.14) the term $\dot{R}$ in terms of decoupling
temperature $T_{D}$ takes the form
\begin{eqnarray*}\dot{R}=2^{-\frac{3}{2}}9(1+\omega)(3\omega-1)\Lambda^{\frac{3}{2}}\mbox{coth}\left(\mbox{arcsinh}\left[\frac{3}{2\pi}\sqrt{\frac{5\Lambda}{2\pi
g_{\ast}G T_{D}^{4}}}\right]\right)\times
\end{eqnarray*}
\begin{equation}
\left(\mbox{sinh}\left(\mbox{arcsinh}\left[\frac{3}{2\pi}\sqrt{\frac{5\Lambda}{2\pi
g_{\ast}G T_{D}^{4}}}\right]\right)^\frac{1}{3}\right)^{-6}
\end{equation}

Correspondingly by substituting $\dot{R}$ from equation (2.16) in
equation (2.6), the baryon-to-entropy ratio in the framework of
Ho$\check{r}$ava-Lifshitz gravity is obtained as
\begin{eqnarray*}
\frac{\eta_{B}}{s}\simeq-\frac{135
g_{b}(1+\omega)(3\omega-1)\Lambda^{\frac{3}{2}}
}{8\sqrt{2}\pi^{2}g_{\ast}M_{\ast}^{2}T_{D}}\mbox{coth}\left[\mbox{arcsinh}\left(\frac{3}{2\pi}\sqrt{\frac{5\Lambda}{2\pi
g_{\ast}G T_{D}^{4}}}\right)\right]\times
\end{eqnarray*}
\begin{equation}
\left[\mbox{sinh}\left(\mbox{arcsinh}\left[\frac{3}{2\pi}\sqrt{\frac{5\Lambda}{2\pi
g_{\ast}G T_{D}^{4}}}\right]\right)^\frac{1}{3}\right]^{-6}
\end{equation}

\begin{figure}
~~~~~~~~~~~~~~~~~~~~~~~~~~\includegraphics[height=1.8in]{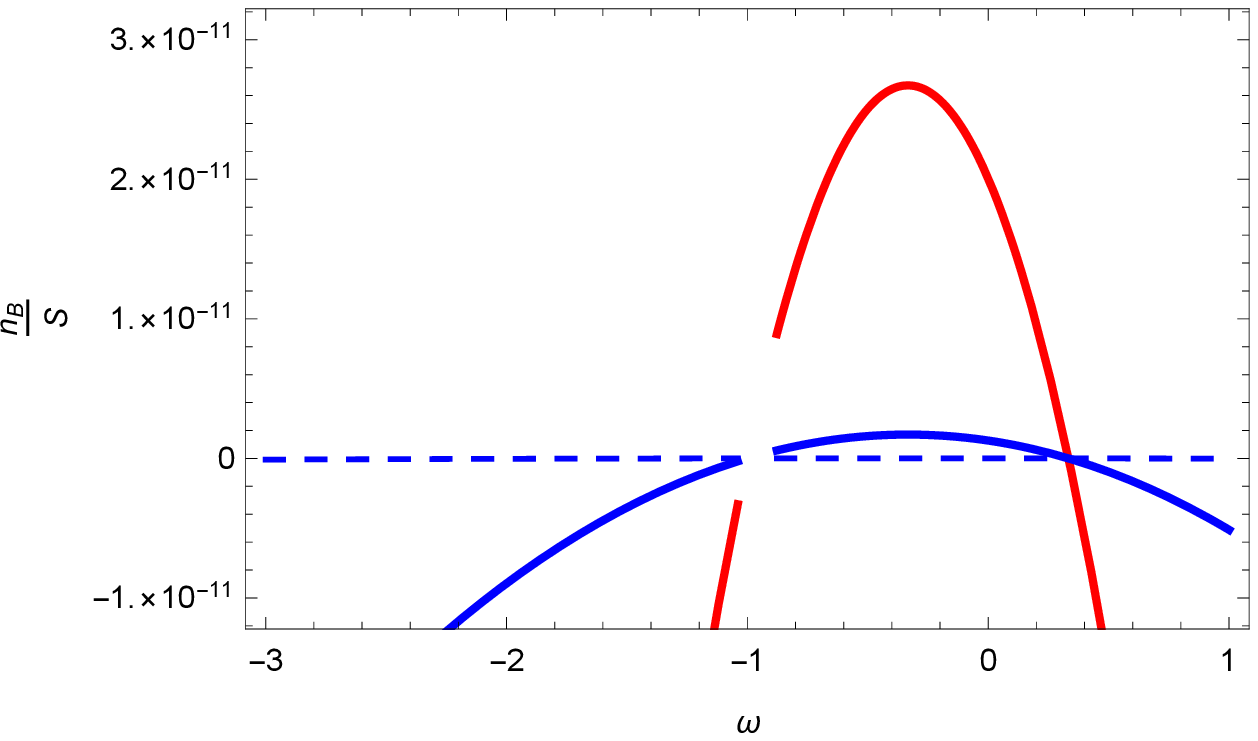}~~~~~~~~~~~~~~~~~

~~~~~~~~~~~~~~~~~~~~~~~~~~~~~~~~~~~~~~~~~~~~~~~~~~~~~~~~~~~~Fig.1~~~~~~~~~~~~~~~~~~~~~~~~\\

\textit{\textbf{Figure 1:} The baryon-to-entropy ratio
$\eta_{B}/s$ is plotted against the EOS parameter $\omega$ for
$\Lambda=10^{45.8}$(dashed curve), $\Lambda=10^{45.5}$(red curve), $\Lambda=10^{45.6}$(blue curve). The parameters are considered as
$-3\leq\omega\leq1.0$, $M_{*}=10^{12}GeV$,
$T_{D}=2\times10^{11}$, $g_{b}\simeq \emph{O}(1)$,
$g_{*}\simeq106$ and $\rho_{0}=10^{102}$.}\vspace{2mm}
\end{figure}

\begin{figure}
\includegraphics[height=1.8in]{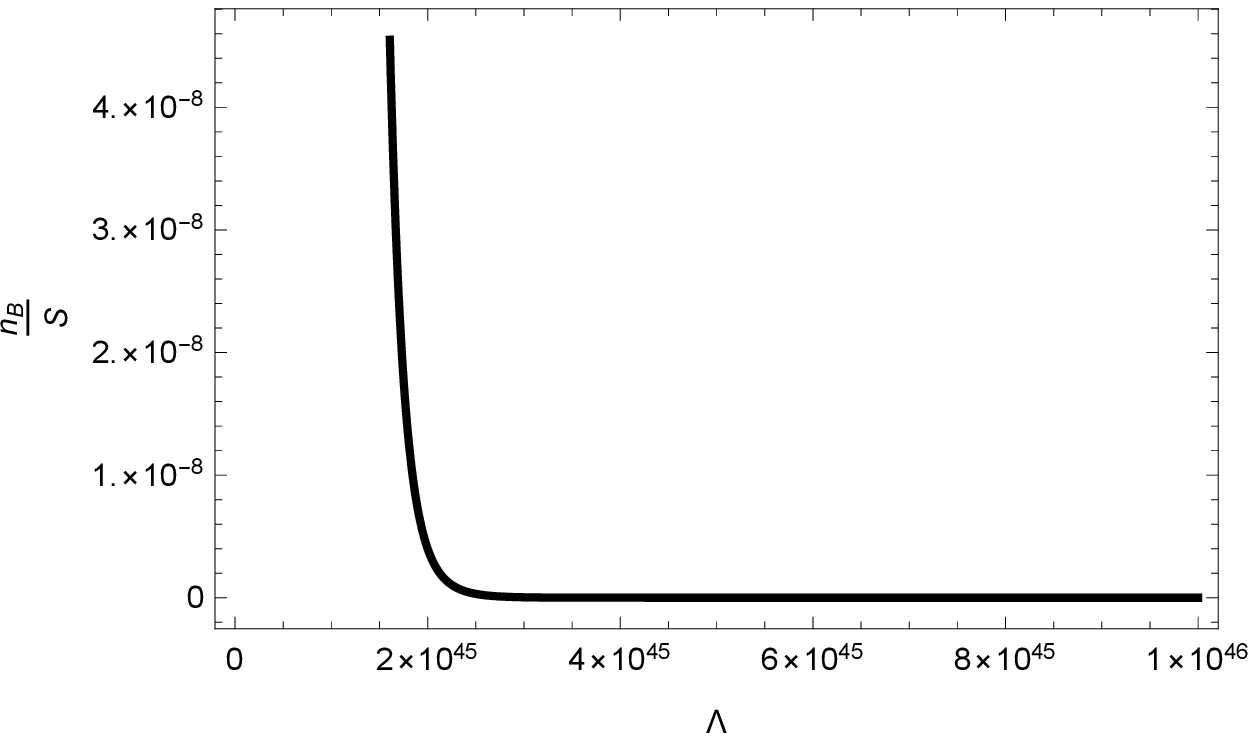}~~~\includegraphics[height=1.8in]{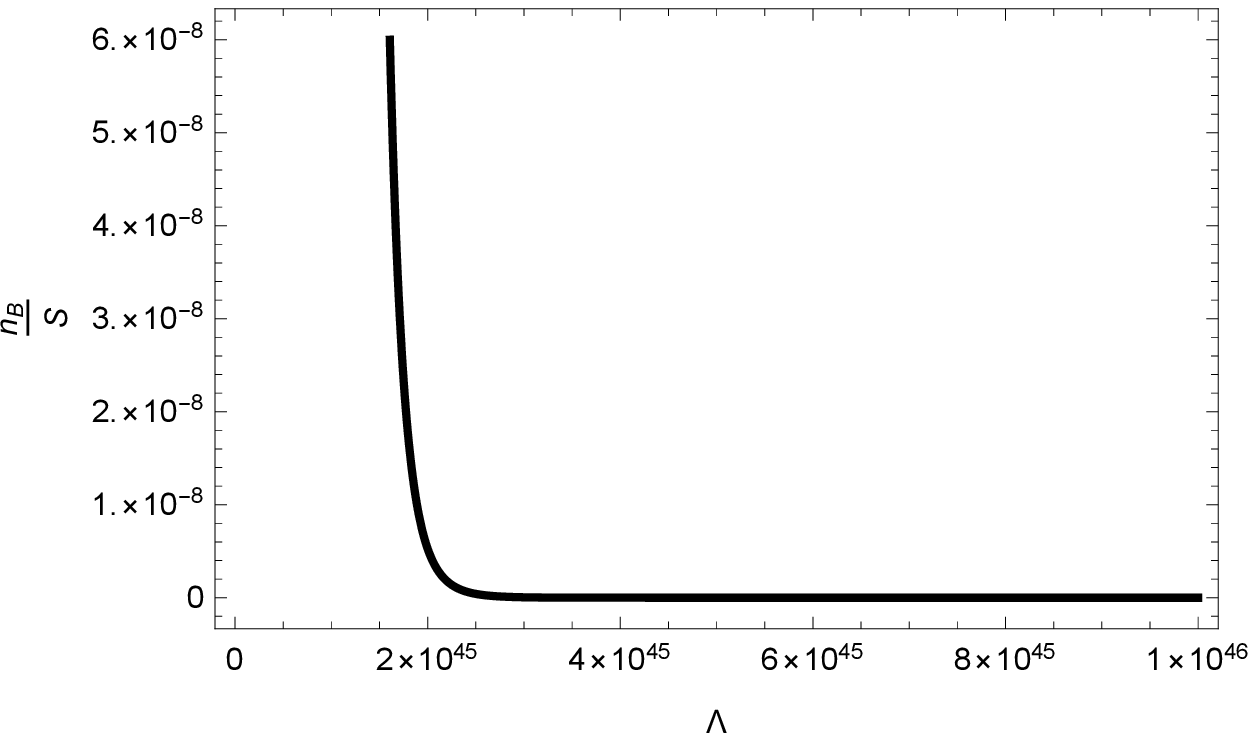}~\\

~~~~~~~~~~~~~~~~~~~~~~~~~~~~~Fig.2(a)~~~~~~~~~~~~~~~~~~~~~~~~~~~~~~~~~~~~~~~~~~~~~~~~~~~~~~Fig.2(b)~~~~~~~~~~~~~~~~\\

\includegraphics[height=1.8in]{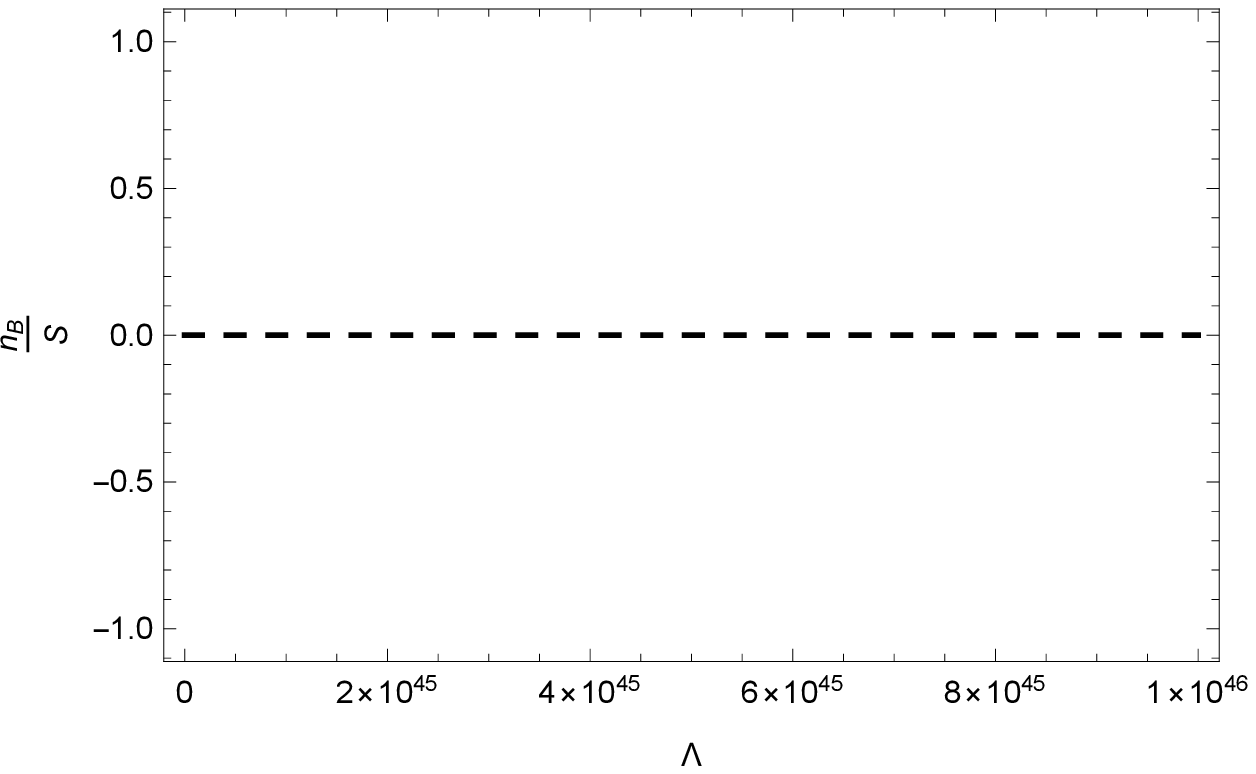}~~~\includegraphics[height=1.8in]{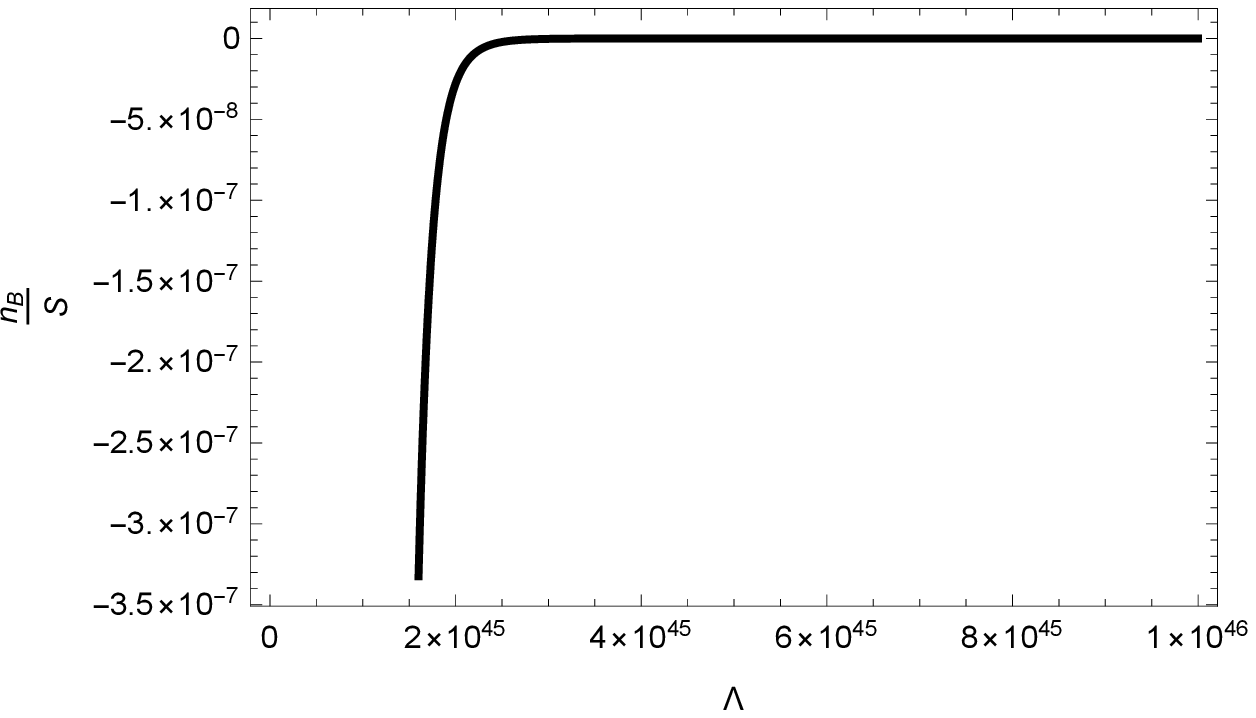}~\\

~~~~~~~~~~~~~~~~~~~~~~~~~~~~~Fig.2(c)~~~~~~~~~~~~~~~~~~~~~~~~~~~~~~~~~~~~~~~~~~~~~~~~~~~~~~Fig.2(d)~~~~~~~~~~~~~~~~\\

~~~~~~~~~~~~~~~~~~~~~~~~~~\includegraphics[height=1.8in]{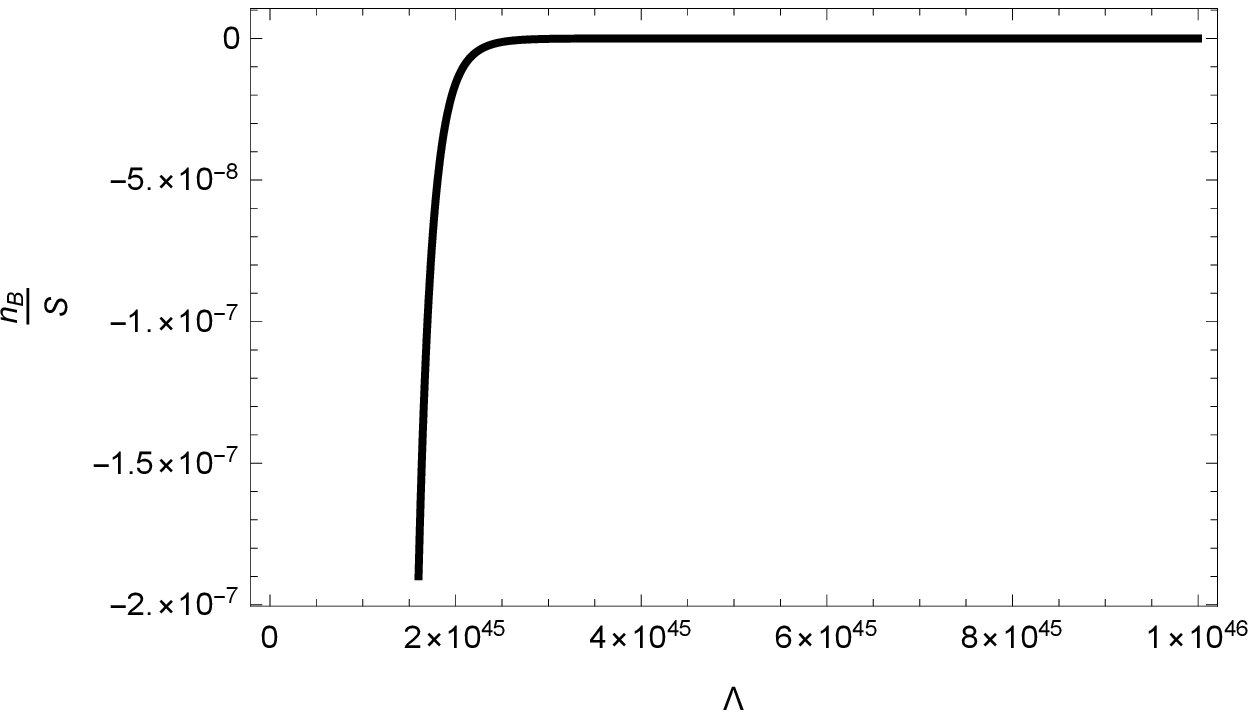}~~~~~~~~~~~~~~~~~\\

~~~~~~~~~~~~~~~~~~~~~~~~~~~~~~~~~~~~~~~~~~~~~~~~~~~~~~~~~~~~Fig.2(e)~~~~~~~~~~~~~~~~~~~~~~~~\\
\textit{\textbf{Figure 2:} The baryon-to-entropy ratio
$\eta_{B}/s$ is plotted against the parameter $\Lambda$ for $\omega=0$ (fig. \mbox{2(a))}, $\omega=-0.3$(fig. \mbox{2(b))}
 $\omega=1/3$(fig. \mbox{2(c))}, $\omega=-2$(fig. \mbox{2(d))} and $\omega=1$(fig. \mbox{2(e))}. The parameters are considered as
$M_{*}=10^{12}$GeV, $T_{D}=2\times10^{11}$,
$g_{b}\simeq \emph{O}(1)$, $g_{*}\simeq106$ and
$\rho_{0}=10^{102}$.}

\end{figure}

 We proceed by investigating those circumstances
under which the resulting baryon-to-entropy ratio can be
compatible with the theoretical bound
$\eta_{B}/s\leq9\times10^{-11}$ \cite{1BCL03}. Here we have
assumed that the cutoff scale $M_{*}=10^{12}$GeV, $g_{b}\simeq \emph{O}(1)$, the total number of the
effectively massless particle in the universe $g_{*}\simeq106$
\cite{1DHKRKGDMHSPJ04,Lambiase1}. We also consider that $\rho_{0}=10^{102}$ and  the decoupling
temperature $T_{D}=2\times10^{11}$. For these values
we have drawn $\eta_{B}/s$ as a function of equation of state
parameter $\omega$ in figure 1 for three different values of the
parameter $\Lambda$. It can be seen that when $-1<\omega<0$, for
$\Lambda=10^{45.5}$ the resulting baryon-to-entropy ratio is
$\eta_{B}/s\leq2.5\times10^{-11}$ and for $\Lambda=10^{45.6}$ ratio is
$\eta_{B}/s\leq2\times10^{-12}$,  which are compatible with the
observational bounds. But for $\omega<-1$, the ratio becomes
negative, therefore this case has no physical interest. The nature
of baryon-to-entropy ratio also crucially depends on the model
parameter $\Lambda$. In fig. 2 we have investigated the nature of
$\eta_{B}/s$ against $\Lambda$ in different cosmological eras
depending on the different values of the EOS parameter $\omega$ by
choosing the previously stated values of the other parameters. The
matter dominated era corresponds to $\omega=0$. We can see from
figure 2 that for $\omega=0$, $\eta_{B}/s\leq5\times10^{-8}$ and
for $\omega=-0.3$, $\eta_{B}/s\leq6\times10^{-8}$ and for both
the cases baryon-to-entropy ratio decrease
as $\Lambda$ gets higher values. So, for the matter dominated era
and for the epoch when the evolution of the universe is driven by
quintessential fluid ($-1<\omega<0$), the baryon-to-entropy ratio
gets the values that are in good agreement with the
observationally accepted value. But for phantom region $\omega<-1$
and for $\omega=1$ the results are physically unacceptable.

\section{Baryogenesis with Scalar Field}
Here we assume that the evolution of the universe is driven as
usual, by a time-varying vacuum expectation value of some scalar
field $\phi$ having energy density and pressure given by
\cite{lucchin}
\begin{equation}
\rho(t)= V(\phi(t))+\frac{1}{2}\dot{\phi(t)}^{2}
\end{equation}
and
\begin{equation}
p(t)= -V(\phi(t))+\frac{1}{2}\dot{\phi(t)}^{2}
\end{equation}
where $V(\phi)$ is the effective potential of the field $\phi$,
the second term represents the kinetic contribution of $\phi$.
Here the ultra-relativistic particle contribution (radiation) has
been neglected. It is assumed that the thermal corrections to the
effective potential are negligible and that the scalar field has
minimal coupling with the geometry \cite{linde}. We also assume
\begin{equation}
\phi(t)=\phi_{0} t^{n} ~~~~ \mbox{and}~~~~ V(\phi(t))=V_{0}
\phi(t)^{m}
\end{equation}
where $n>1$ and $m$ are positive constants. With this assumption
(3.1) becomes
\begin{equation}
\rho(t)=\frac{1}{2} n^{2}
t^{2n-2}\phi_{0}^{2}+V_{0}t^{nm}\phi_{0}^{m}
\end{equation}
Using equation (3.4) in equation (2.2) for flat universe we get
the scale factor as
\begin{equation}
a(t)=C_{2}\mbox{exp}\left[t\sqrt{\frac{\Lambda}{2}}+\frac{4\pi
G}{3\sqrt{2\Lambda}}\left(\frac{n^{2}t^{2n-1}\phi_{0}^{2}}{2n-1}+\frac{2
V_{0}\phi_{0}^{m}t^{mn+1}}{1+mn}\right)\right]
\end{equation}
where $C_{2}$ is an arbitrary integration constant. From equation
(2.7) and equation (3.4) we have obtained the relation between
decoupling time $t_{D}$ and critical temperature $T_{D}$ for $m=1$
and $n=2$ as
\begin{equation}
t_{D}=\frac{\sqrt{g}\pi
T_{D}^{2}}{\sqrt{30\phi_{0}(V_{0}+2\phi_{0})}}
\end{equation}
Using equation (3.5) for $m=1,~n=2$ in (2.10) we get $\dot{R}$ for
flat universe
\begin{equation}
\dot{R}=16\pi Gt\phi_{0}(1-3\omega)(V_{0}+2\phi_{0})
\end{equation}
Exploiting equation (3.6), (3.7) becomes
\begin{equation}
\dot{R}=8\sqrt{\frac{2 g}{15}}\pi^{2} G
T_{D}^{2}(3\omega-1)\sqrt{\phi_{0}(V_{0}+2\phi_{0})}
\end{equation}
With the help of equations (3.8) and (2.6) the baryon-to entropy
ratio in this scenario reads
\begin{equation}
\frac{\eta_{B}}{s}\simeq\frac{2\sqrt{30}G
g_{b}T_{D}(1-3\omega)\sqrt{\phi_{0}(V_{0}+2\phi_{0})}}{\sqrt{g}M^{2}}
\end{equation}
\begin{figure}
~~~~~~~~~~~~~~~~~~~~~~~~~~\includegraphics[height=1.8in]{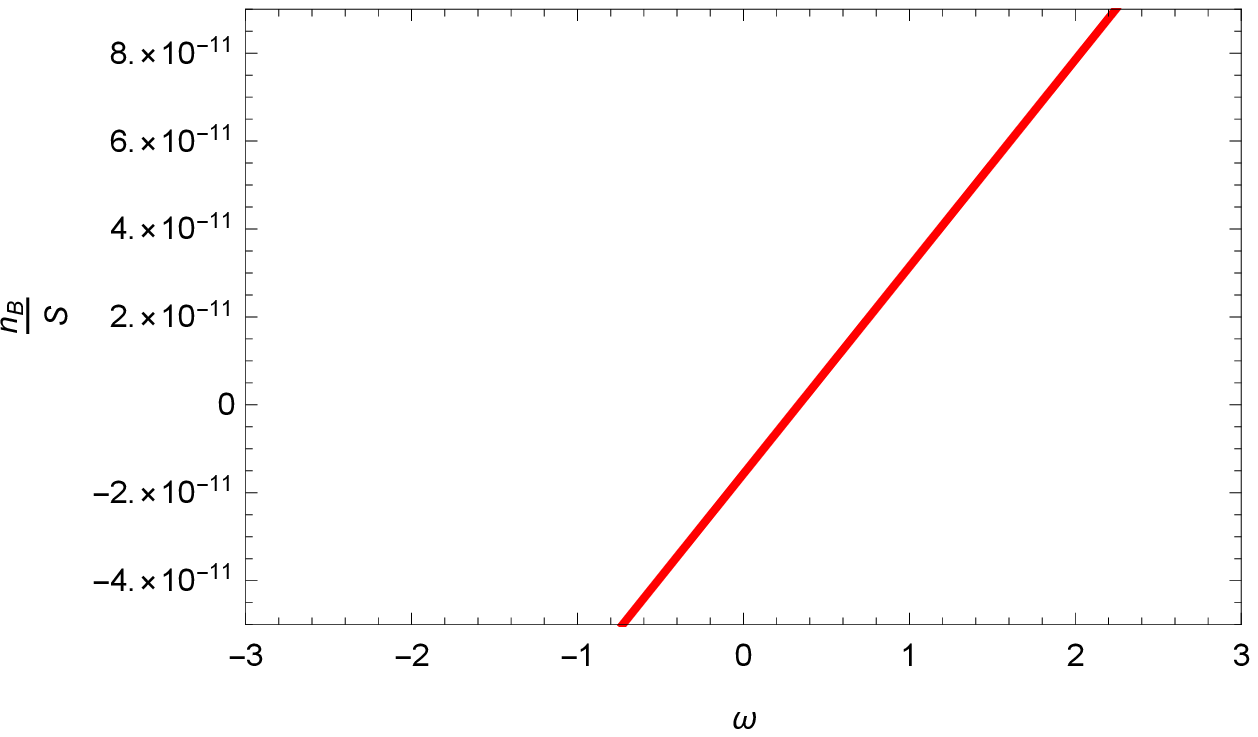}~~~~~~~~~~~~~~~~~

~~~~~~~~~~~~~~~~~~~~~~~~~~~~~~~~~~~~~~~~~~~~~~~~~~~~~~~~~~~~Fig.3~~~~~~~~~~~~~~~~~~~~~~~~\\

\textit{\textbf{Figure 3:} The baryon - to - entropy ratio
$\eta_ {B}/s$ for scalar field is plotted against the EOS parameter $\omega$.The parameters are considered as
$ - 3\leq\omega\leq3 $, $M_ {*} = 10^{12} GeV$, $T_ {D}=2\times10^{16} $, $g_ {b}\simeq \emph{O}(1) $, $g_ {*}\simeq106$, $\phi_{0}=10^{7}$ \mbox{and} $V_{0}=10^{5}$.}\vspace{2mm}
\end{figure}
\begin{figure}
\includegraphics[height=1.8in]{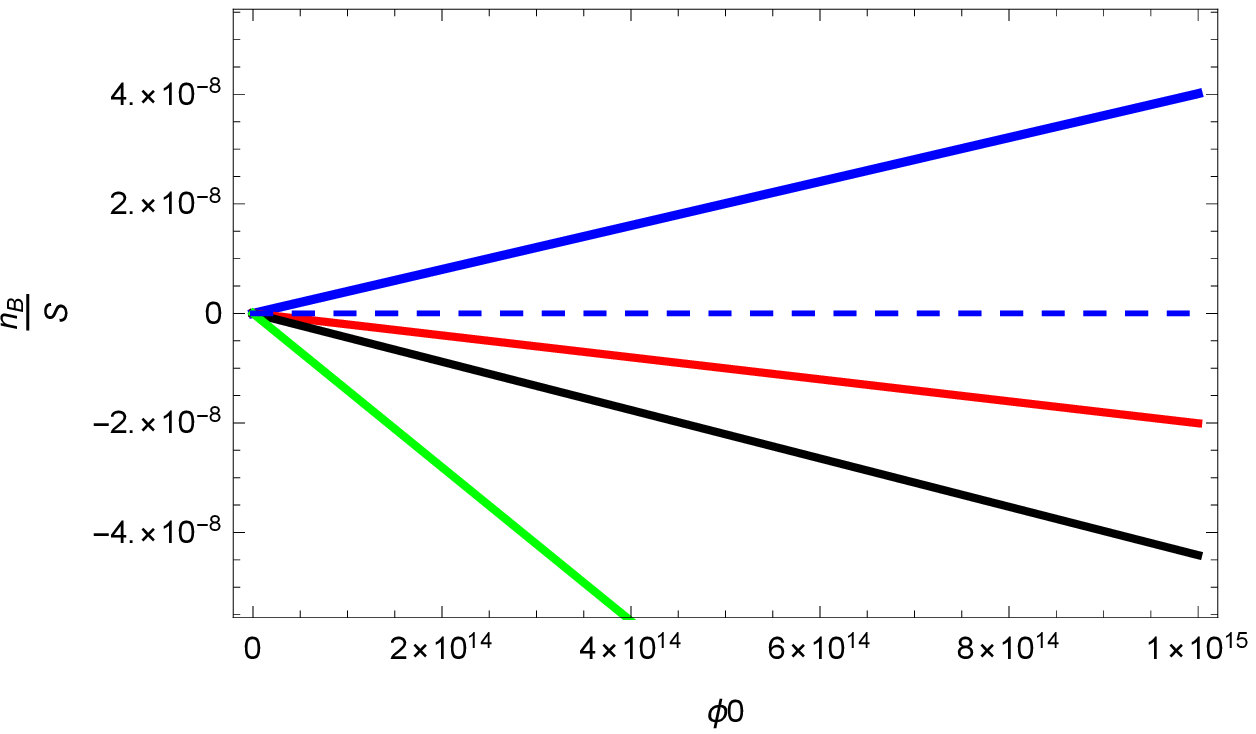}~~~\includegraphics[height=1.8in]{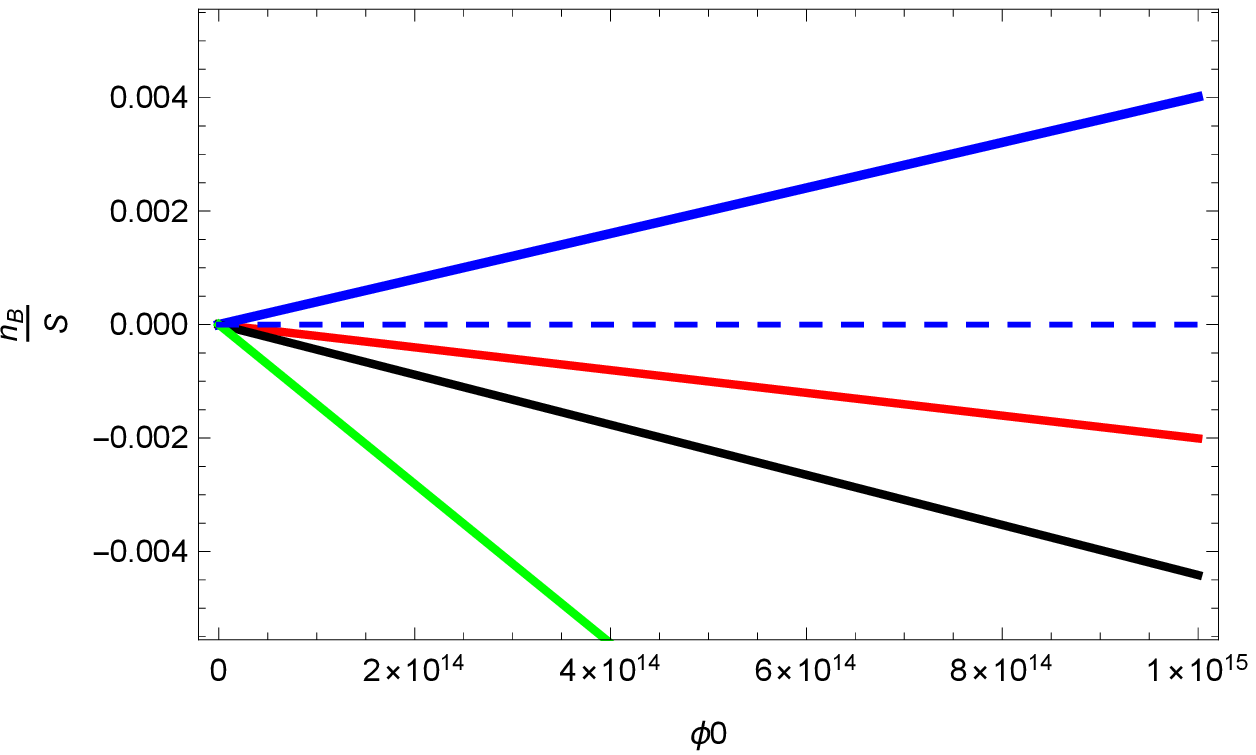}~\\

~~~~~~~~~~~~~~~~~~~~~~~~~~~~~Fig.4(a)~~~~~~~~~~~~~~~~~~~~~~~~~~~~~~~~~~~~~~~~~~~~~~~~~~~~~~Fig.4(b)~~~~~~~~~~~~~~~~\\

\textit{\textbf{Figure 4:} The baryon-to-entropy ratio
$\eta_{B}/s$ for scalar field is plotted against the parameter $\phi_{0}$ for $\omega=-2$(green curve),$\omega=-0.4$(black curve),
$\omega=0$ (red curve), $\omega=1/3$ (dashed blue curve) and $\omega=1$(blue curve). The parameters are considered as $M_{*}=10^{12}GeV$,
 $g_ {b}\simeq \emph{O}(1) $,
$g_{*}\simeq106$ and $V_{0}=10^{10}$. The left plot corresponds to $T_{D}=2\times10^{11}$ while the right plot corresponds to $T_{D}=2\times10^{16}$.}\vspace{2mm}
\end{figure}
\begin{figure}
\includegraphics[height=1.8in]{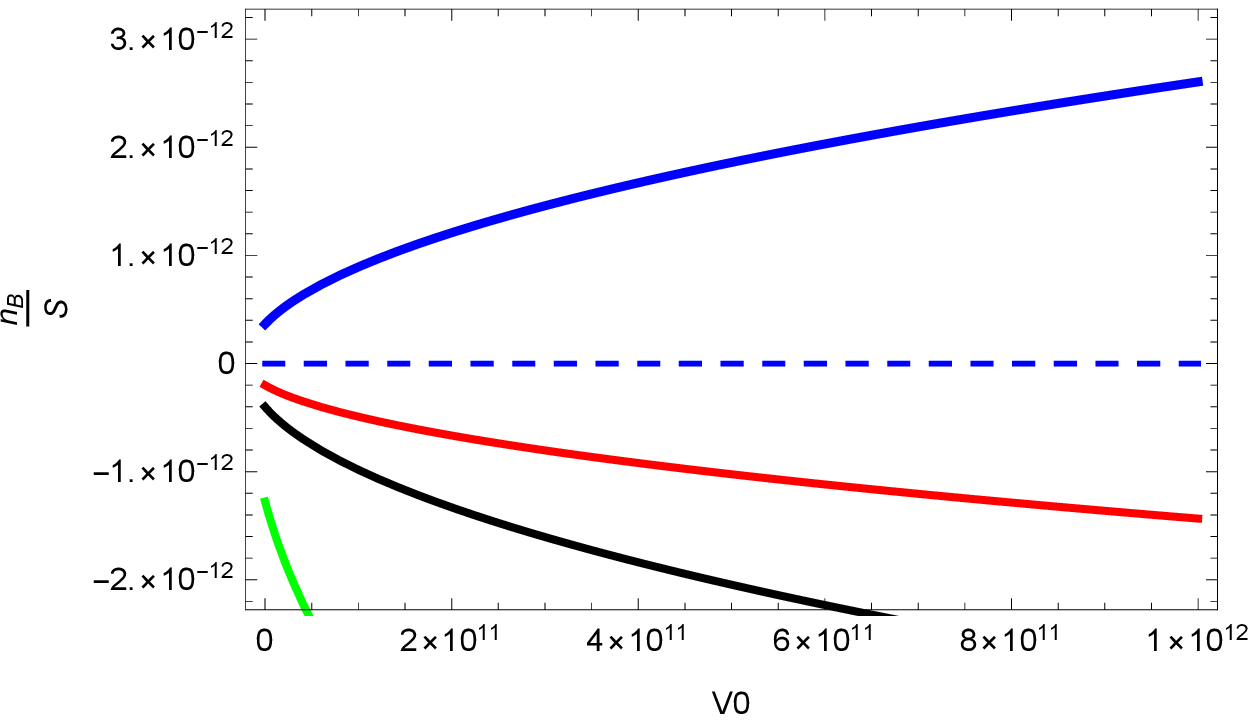}~~~\includegraphics[height=1.8in]{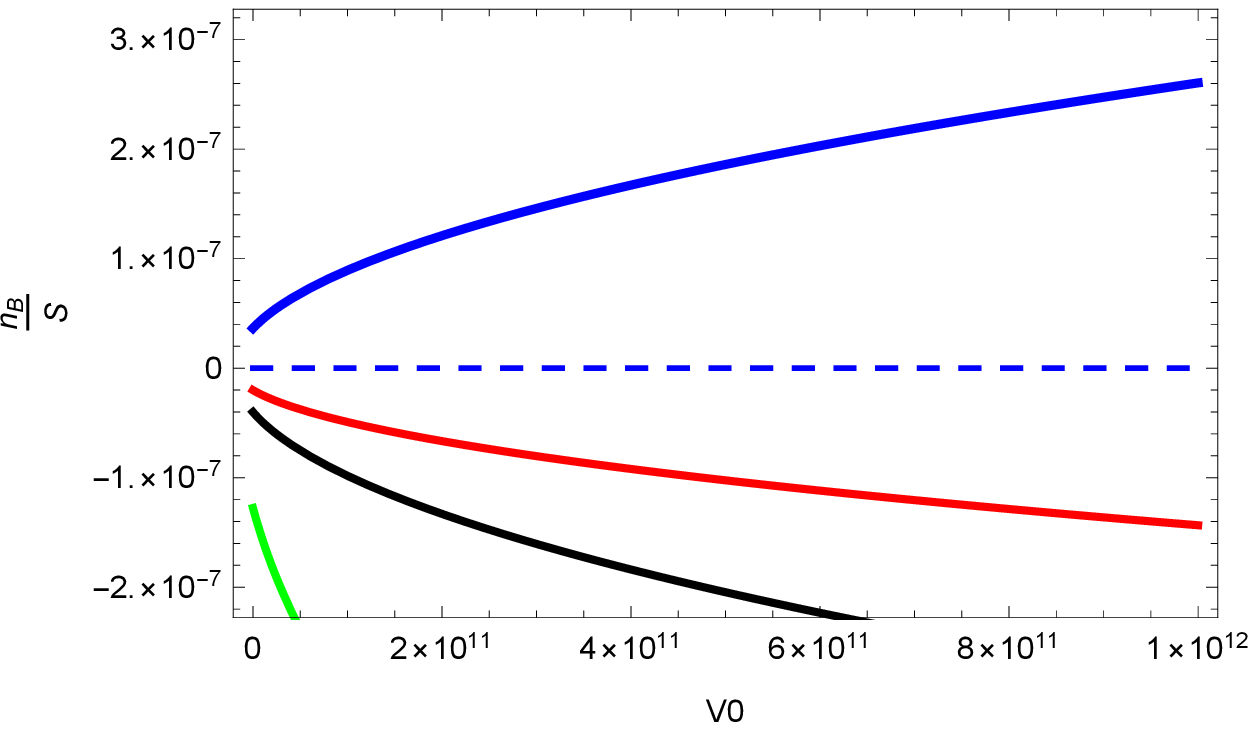}~\\

~~~~~~~~~~~~~~~~~~~~~~~~~~~~~Fig.5(a)~~~~~~~~~~~~~~~~~~~~~~~~~~~~~~~~~~~~~~~~~~~~~~~~~~~~~~Fig.5(b)~~~~~~~~~~~~~~~~\\
\textit{\textbf{Figure 5:} The baryon-to-entropy ratio
$\eta_{B}/s$ for scalar field is plotted against the parameter $V_{0}$ for $\omega=-2$(green curve),$\omega=-0.4$(black curve),
$\omega=0$ (red curve), $\omega=1/3$ (dashed blue curve) and $\omega=1$(blue curve). The parameters are considered as
$M_{*}=10^{12}$GeV,
$g_ {b}\simeq \emph{O}(1) $, $g_{*}\simeq106$ and
$\phi_{0}=10^{10}$.The left plot corresponds to $T_{D}=2\times10^{11}$ while the right plot corresponds to $T_{D}=2\times10^{16}$.}
\end{figure}
In figure 3 we have investigated the profile of baryon to entropy
ratio for scalar field against the EOS parameter $\omega$ for
$\phi_{0}=10^{7}$ and $V_{0}=10^{5}$. In this scenario we have
found $\eta_{B}/s\leq8\times10^{-11}$. In figure 4(a) and figure 4(b) we
have plotted the behavior of the baryon-to-entropy ratio for
scalar field as a function of $\phi_{0}$ for $T_{D}=2\times10^{11}$ and $T_{D}=2\times10^{16}$ respectively
with the previously described values of the parameters $M_{*}$,
 $g_{b}$, $g_{*}$ and $V_{0}=10^{10}$. It is observed that $\eta_{B}/s$ takes observationally accepted value as the decoupling temperature $T_{D}$ attains lower values. For $T_{D}=2\times10^{11}$ the baryon-to entropy ratio becomes $\eta_{B}/s\leq4.5\times10^{-8}$. But for $T_{D}=2\times10^{16}$, the baryon-to entropy ratio gets value $\eta_{B}/s\leq4\times10^{-3}$ which is far from the observational bound. In each case, $\eta_{B}/s$ increases as $\phi_{0}$ increases for $\omega=1$. In figure 5(a) and 5(b) we have plotted $\eta_{B}/s$ for
scalar field as a function of $V_{0}$ for $T_{D}=2\times10^{11}$ and $T_{D}=2\times10^{16}$ respectively and have
find that $\eta_{B}/s$ increases as $V_{0}$ increases for
$\omega=1$. In this era,the baryon-to entropy ratio becomes $\eta_{B}/s\leq3\times10^{-12}$ for $T_{D}=2\times10^{11}$ and $\eta_{B}/s\leq3\times10^{-7}$ for $T_{D}=2\times10^{16}$ . For the other values of $\omega $ we get un-physical results.

\section{Baryogenesis with Generalized Chaplygin Gas}
As discussed earlier the energy density of Generalized Chaplygin Gas (GCG) \cite{1BMCBOSAA02} is given by $p_{GCG}=-\frac{A}{\rho_{GCG}^{\alpha}}$, where $0<\alpha\leq 1$ and A is a positive constant.
In the framework of FRW cosmology, with the energy conservation equation it leads to
\begin{equation}
\rho_{GCG}=\left(A+B a^{-3(1+\alpha)}\right)^{\frac{1}{1+\alpha}}
\end{equation}
where $0<\alpha\leq 1$, A is a positive constant and B is a
positive integration constant. Using equation (2.2) and (4.1) we
obtain the scale factor for a flat universe as follows
\begin{equation}
a(t)=\left[\frac{e^{\frac{(1+\alpha)}{2\sqrt{\Lambda}}(3\Lambda +8
G \pi
A^{\frac{1}{1+\alpha}})(\sqrt{2}t+6\sqrt{\Lambda}C_{3}(1+\alpha))}-8\pi
G B} {(1+\alpha)(3\Lambda +8 G \pi
A^{\frac{1}{1+\alpha}})}\right]^{\frac{1}{3(1+\alpha)}}
\end{equation}
where $C_{3}$ is an integration constant. The above choice of scale factor yields the Hubble parameter as
\begin{equation}
H(t)=\left[\frac{(3\Lambda +8 G \pi
A^{\frac{1}{1+\alpha}})e^{\frac{(1+\alpha)}{2\sqrt{\Lambda}}(3\Lambda
+8 G \pi
A^{\frac{1}{1+\alpha}})(\sqrt{2}t+6\sqrt{\Lambda}C_{3}(1+\alpha))}}{3\sqrt{2\Lambda}\left(e^{\frac{(1+\alpha)}
{2\sqrt{\Lambda}}(3\Lambda +8 G \pi
A^{\frac{1}{1+\alpha}})(\sqrt{2}t+6\sqrt{\Lambda}C_{3}(1+\alpha))}-8\pi
G B\right)}\right]
\end{equation}
Correspondingly the energy density in terms of cosmic time becomes
\begin{equation}
\rho_{GCG}(t)=\left(A+\frac{B(1+\alpha)(3\Lambda +8 G \pi
A^{\frac{1}{1+\alpha}})}{e^{\frac{(1+\alpha)}
{2\sqrt{\Lambda}}(3\Lambda +8 G \pi
A^{\frac{1}{1+\alpha}})(\sqrt{2}t+6\sqrt{\Lambda}C_{3}(1+\alpha))}-8\pi
G B} \right)^{\frac{1}{1+\alpha}}
\end{equation}
Combining equations (4.4) and (2.7) we get the relation between
decoupling time $t_{D}$ and critical temperature $T_{D}$ in this
scenario, which is given by

\begin{eqnarray*}
t_{D}=\frac{B}{\sqrt{2}(1+\alpha)(3\Lambda+8 G \pi
A^{\frac{1}{1+\alpha}})}\left[8\pi G
+\frac{(1+\alpha)30^{(1+\alpha)}(3\Lambda+8 G \pi
A^{\frac{1}{1+\alpha}})}{\pi^{2(1+\alpha)}g^{(1+\alpha)}T_{D}^{4(1+\alpha)}-30^{(1+\alpha)A}}\right]
\end{eqnarray*}
\begin{equation}
-6(1+\alpha)C_{3}\sqrt{\frac{\Lambda}{2}}
\end{equation}

\begin{figure}
\includegraphics[height=1.8in]{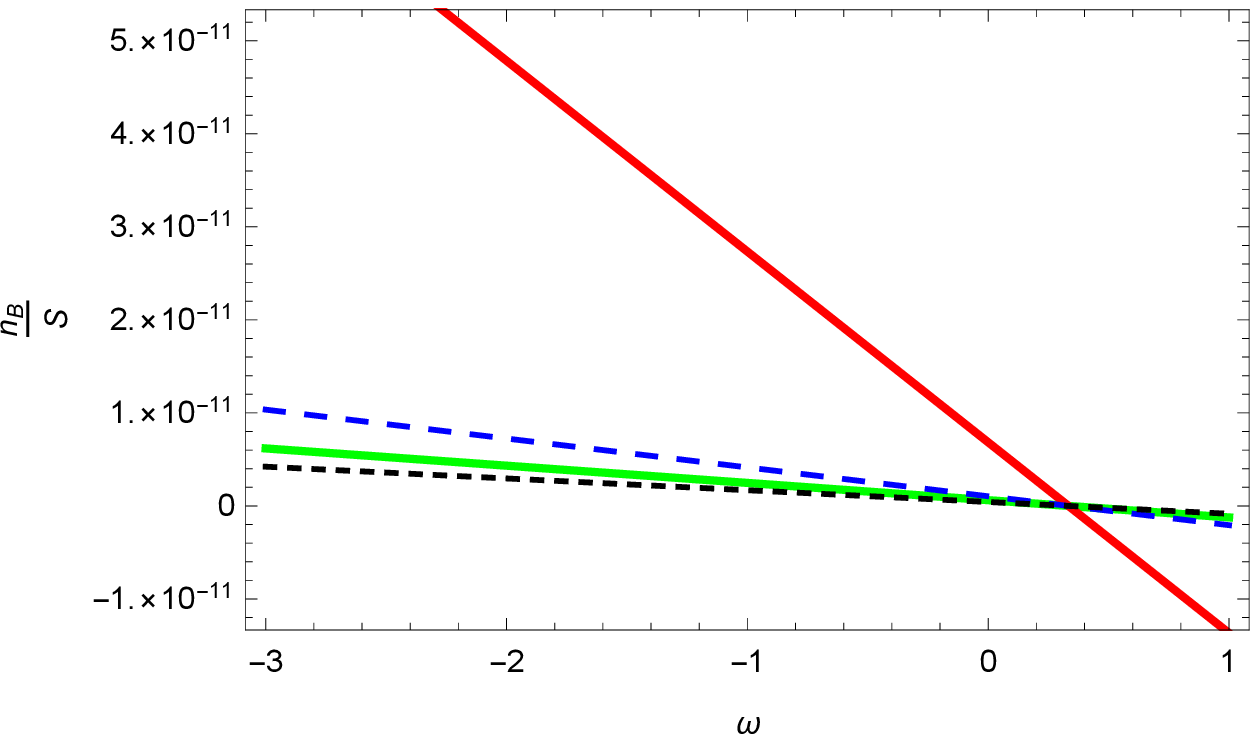}~~~\includegraphics[height=1.8in]{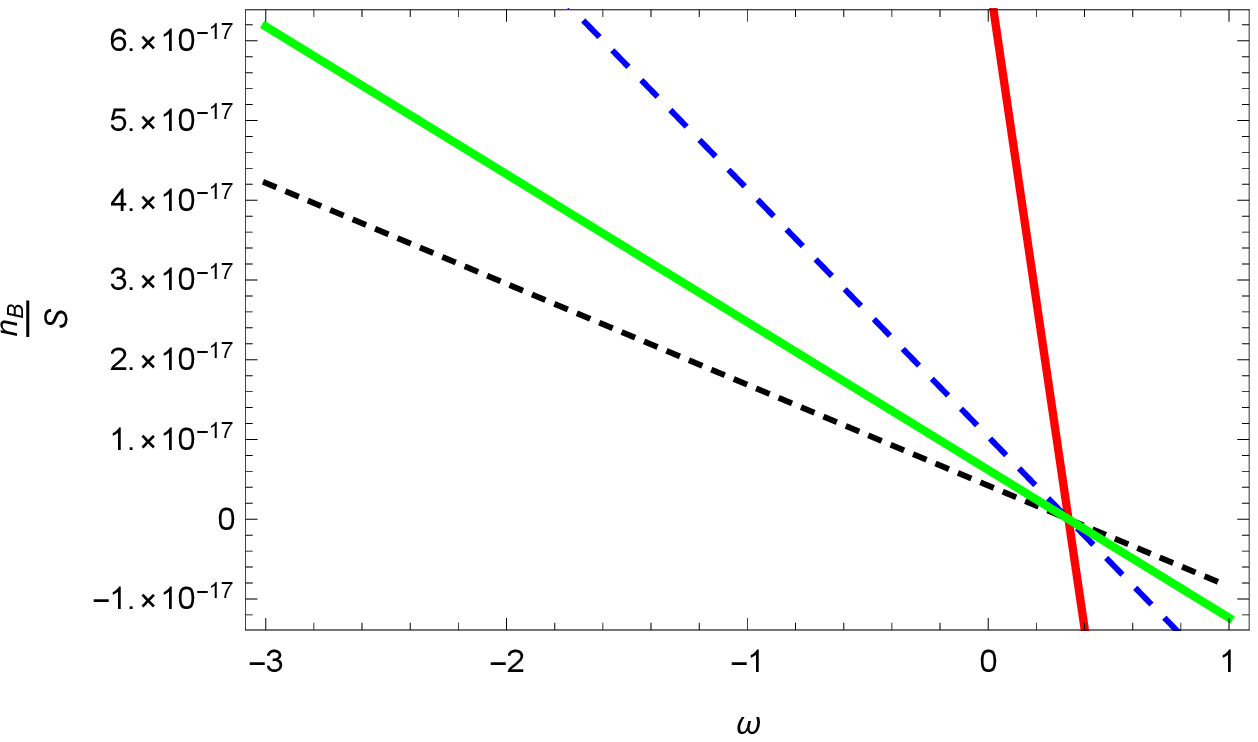}~\\

~~~~~~~~~~~~~~~~~~~~~~~~~~~~~Fig.6(a)~~~~~~~~~~~~~~~~~~~~~~~~~~~~~~~~~~~~~~~~~~~~~~~~~~~~~~Fig.6(b)~~~~~~~~~~~~~~~~\\

\textit{\textbf{Figure 6:} The baryon - to - entropy ratio
$\eta_ {B}/s$ for GCG is plotted against the EOS parameter $\omega$ for
$\alpha = 0.8$ (red curve), $\alpha =0.93$ (dashed blue curve), $\alpha =1$ (dashed black curve), $\alpha =0.97$ (green curve).The parameters are considered as
$ - 3\leq\omega\leq1 .0 $, $M_ {*} = 10^{12} GeV$, $g_ {b}\simeq \emph{O}(1)$, $g_ {*}\simeq106$, $A=5.5\times10^{10}$ and $B=9\times10^{-25}$. Fig. 6(a) corresponds to  $T_ {D}=2\times10^{11} $ and Fig. 6(b)corresponds to  $T_ {D}=2\times10^{16} $ }\vspace{2mm}
\end{figure}
From equations (4.4) and (2.10) we get
\begin{eqnarray*}
\dot{R}=-4\sqrt{\frac{2}{\Lambda}}G \pi
B(1-3\omega)(1+\alpha)e^{\frac{(1+\alpha)}{2\sqrt{\Lambda}}(3\Lambda
+8 G \pi A^{\frac{1}{1+\alpha}})
(\sqrt{2}t+6\sqrt{\Lambda}C_{3}(1+\alpha))}\times
\end{eqnarray*}
\begin{eqnarray*}
(3\Lambda +8 G \pi
A^{\frac{1}{1+\alpha}})^{2}\left(e^{\frac{(1+\alpha)}{2\sqrt{\Lambda}}(3\Lambda
+8 G \pi
A^{\frac{1}{1+\alpha}})(\sqrt{2}t+6\sqrt{\Lambda}C_{3}(1+\alpha))}
-8\pi G B\right)^{-2}\times
\end{eqnarray*}
\begin{equation}
\left(A+\frac{B(1+\alpha)(3\Lambda +8 G \pi
A^{\frac{1}{1+\alpha}})}{e^{\frac{(1+\alpha)}{2\sqrt{\Lambda}}(3\Lambda
+8 G \pi
A^{\frac{1}{1+\alpha}})(\sqrt{2}t+6\sqrt{\Lambda}C_{3}(1+\alpha))}-8\pi
G B}\right)^{-1+\frac{1}{1+\alpha}}
\end{equation}

Again using equation (4.5) the above form of $\dot{R}$ is
expressed in terms of $T_{D}$
\begin{eqnarray*}
\dot{R}=-4\sqrt{\frac{2}{\Lambda}}G \pi
B(1-3\omega)(1+\alpha)e^{\frac{B}{\sqrt{\Lambda}}\left(4 \pi
G+\frac{2^{\alpha}15^{(1+\alpha)}(1+\alpha) (3\Lambda +8 G \pi
A^{\frac{1}{1+\alpha}})}{g^{(1+\alpha)}\pi^{2(1+\alpha)}T_{D}^{4(1+\alpha)}-A30^{(1+\alpha)}}\right)}\times
\end{eqnarray*}
\begin{eqnarray*}
(3\Lambda +8 G \pi
A^{\frac{1}{1+\alpha}})^{2}\left(e^{\frac{B}{\sqrt{\Lambda}}\left(4
\pi G+\frac{2^{\alpha}15^{(1+\alpha)}(1+\alpha)(3\Lambda +8 G \pi
A^{\frac{1}{1+\alpha}})}
{g^{(1+\alpha)}\pi^{2(1+\alpha)}T_{D}^{4(1+\alpha)}-A30^{(1+\alpha)}}\right)}-8\pi
G B\right)^{-2}\times
\end{eqnarray*}
 \begin{equation}
\left(A+\frac{B(1+\alpha)(3\Lambda +8 G \pi
A^{\frac{1}{1+\alpha}})}{e^{\frac{B}{\sqrt{\Lambda}}\left(4 \pi
G+\frac{2^{\alpha}15^{(1+\alpha)}(1+\alpha)(3\Lambda +8 G \pi
A^{\frac{1}{1+\alpha}})}{g^{(1+\alpha)}\pi^{2(1+\alpha)}T_{D}^{4(1+\alpha)}-A30^{(1+\alpha)}}\right)}-8\pi
G B}\right)^{-1+\frac{1}{1+\alpha}}
\end{equation}

\begin{figure}
~~~~~~~~~~~~~~~~~~~~~~~~~~\includegraphics[height=1.8in]{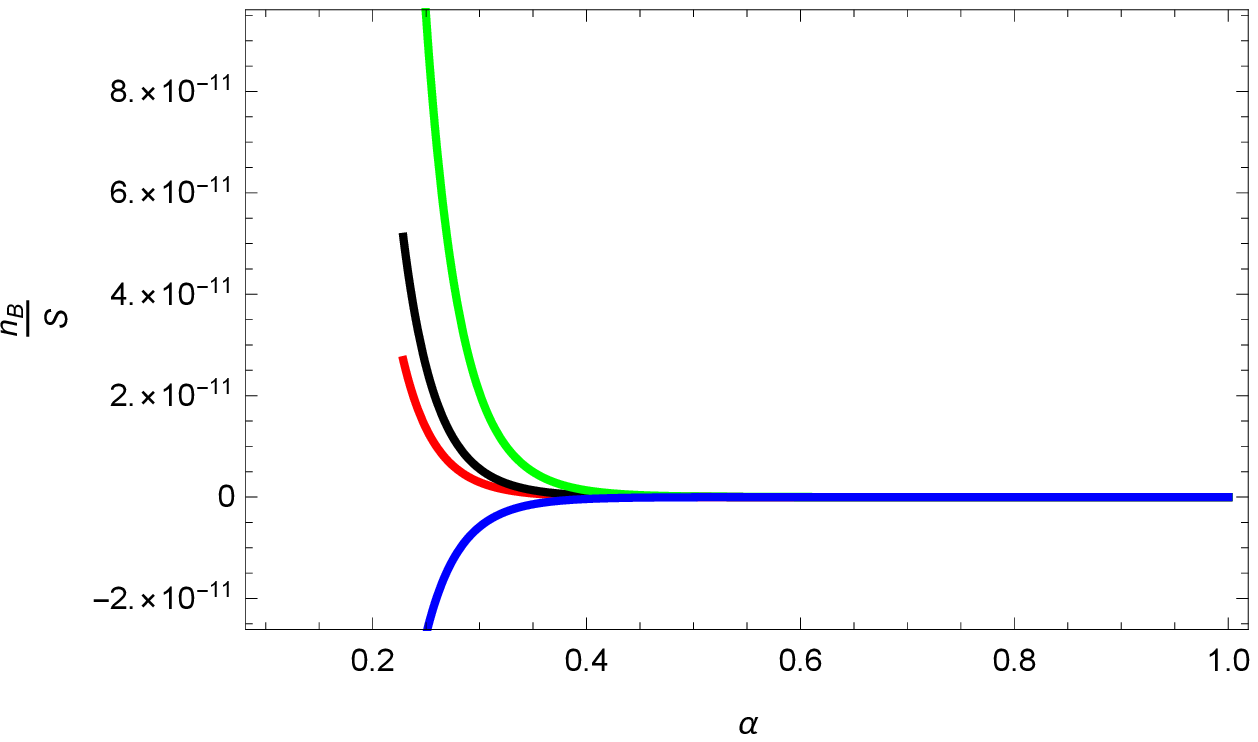}~~~~~~~~~~~~~~~~~

~~~~~~~~~~~~~~~~~~~~~~~~~~~~~~~~~~~~~~~~~~~~~~~~~~~~~~~~~~~~Fig.7~~~~~~~~~~~~~~~~~~~~~~~~\\
\vspace{4mm}

\textit{\textbf{Figure 7:} The baryon-to-entropy ratio
$\eta_{B}/s$ for GCG is plotted against the GCG model parameter $\alpha$ for $\omega=-2$(green curve),$\omega=-0.3$(black curve),
$\omega=0$ (red curve) and $\omega=1$(blue curve). The parameters are considered as $M_{*}=10^{12}GeV$,
$T_{D}=2\times10^{16}$, $g_{b}\simeq \emph{O}(1)$,
$g_{*}\simeq106$, $A=5.5\times10^{10}$, $B=9\times10^{-25}$ and $\Lambda=10^{45.5}$.}\vspace{2mm}

\end{figure}
Inserting equation (4.7) into (2.6) the baryon-to entropy reads

\begin{eqnarray*}
\frac{\eta_{B}}{s}\simeq \frac{15\sqrt{2}B G
g_{b}(1-3\omega)(1+\alpha)}{\pi g T_{D}
M_{\ast}\sqrt{\Lambda}}e^{\frac{B}{\sqrt{\Lambda}}\left(4 \pi
G+\frac{2^{\alpha}15^{(1+\alpha)} (1+\alpha)(3\Lambda +8 G \pi
A^{\frac{1}{1+\alpha}})}{g^{(1+\alpha)}\pi^{2(1+\alpha)}T_{D}^{4(1+\alpha)}-A30^{(1+\alpha)}}\right)}\times
\end{eqnarray*}
\begin{eqnarray*}
(3\Lambda +8 G \pi A^{\frac{1}{1+\alpha}})^{2}
\left(e^{\frac{B}{\sqrt{\Lambda}}\left(4 \pi
G+\frac{2^{\alpha}15^{(1+\alpha)}(1+\alpha)(3\Lambda +8 G \pi
A^{\frac{1}{1+\alpha}})}{g^{(1+\alpha)}\pi^{2(1+\alpha)}T_{D}^{4(1+\alpha)}-A30^{(1+\alpha)}}\right)}-8\pi
G B\right)^{-2}\times
\end{eqnarray*}
\begin{equation}
\left(A+\frac{B(1+\alpha)(3\Lambda +8 G \pi
A^{\frac{1}{1+\alpha}})}{e^{\frac{B}{\sqrt{\Lambda}} \left(4 \pi
G+\frac{2^{\alpha}15^{(1+\alpha)}(1+\alpha)(3\Lambda +8 G \pi
A^{\frac{1}{1+\alpha}})}{g^{(1+\alpha)}\pi^{2(1+\alpha)}T_{D}^{4(1+\alpha)}
-A30^{(1+\alpha)}}\right)}-8\pi G
B}\right)^{-1+\frac{1}{1+\alpha}}
\end{equation}

By choosing $M_ {*} = 10^{12} GeV$, $g_ {b}\simeq \emph{O}(1)$,
$g_ {*}\simeq106$, $A=5.5\times10^{10}$ and $B=9\times10^{-25}$ we
have plotted the functional dependence of baryon-to-entropy ratio
as a function of EOS parameter $\omega$ in figure 6 for four
different values of the GCG parameter $\alpha$. It can be seen
that corresponding to $T_ {D}=2\times10^{11}$, baryon-to-entropy
ratio predicted by (4.8) is $\eta_{B}/s\leq 8.5\times10^{-11}$ for
$\alpha=0.8$, which is very close to observationally accepted
value. Where as the predicted value of baryon-to-entropy ratio
corresponding to $T_ {D}=2\times10^{16} $ is $\eta_{B}/s\leq
7\times10^{-16}$ for $\alpha=0.8$. It is also observed that as
$\alpha$ takes higher values between $0.8<\alpha<1$ the ratio
assumes smaller values compared to the observed value. Thus the
model parameter $\alpha$ affects the ratio in a crucial way.
 Next in
figure 7, we have investigated the nature of the baryon-to-entropy
ratio as a function of GCG parameter $\alpha$ in various phase of
the evolution of the universe. It  can be seen that
in the matter dominated era, for quintessential fluid and for
phantum dark energy baryon-to-entropy ratio takes positive values
and $\eta_{B}/s$ decreases as $\alpha$ increases. But for
$\omega=1$, $\eta_{B}/s$ shows un-physical result.

\section{\normalsize\bf{Conclusion}}
Here we have investigated the gravitational baryogenesis mechanism
under the framework of Ho$\check{r}$ava-Lifshitz gravity. The
baryon-to-entropy ratio is computed for Ho$\check{r}$ava-Lifshitz
gravity in terms of scale factor and finally in terms of the
decoupling temperature.

In contrary to the other works done on the gravitational
baryogenesis \cite{Lambiase1,Odintsov1,1OSDOVK16}, our computed
Baryon-to-entropy ratio in the framework of
Ho$\check{r}$ava-Lifshitz gravity is independent of the critical
density parameter $\rho_{0}$. Baryon-to-entropy ratio have been
generated against the EOS parameter $\omega$ and the model
parameter $\Lambda$. We have seen that these parameters play a
crucial role to make the value compatible with the observational
bound. It should be noted that to make the considered scenario viable, we have assumed the decoupling temperature $T_{D}=2\times10^{11}$. With this assumption we have observed that if the expansion of the universe is
driven by quintessential fluid the ratio becomes consistent with
the observational bounds for $\Lambda=10^{45.5}$ and
$\Lambda=10^{45.6}$. Also in case of matter dominated universe we
have seen that the ratio gets values compatible with the
observational data for these two values of $\Lambda$. But when the
evolution of the universe is dominated by phantom like dark
energy, we get un-physical results. In the radiation dominated
epoch ($\omega=1/3$) the baryon-to-entropy ratio becomes zero like
Einstein-Hilbert case. In the early universe prior to radiation
dominated epoch ($\omega>1/3$) we get un-physical results.

Further we have investigated the baryogenesis mechanism for a
scalar field in the context of Ho$\check{r}$ava-Lifshitz gravity.
We have chosen the scalar field and the scalar potential in the
power-law form and have computed the scale factor in a general
case. Then we have investigated the mechanism for particular
values of the power-law parameters. For this selection, we have
studied the behavior of the baryon-to-entropy ratio as a function
of EoS parameter $\omega$ for $T_{D}=2\times10^{16}$ . We have
seen that the ratio $\eta_{B}/s$ always increases as $\omega$
increases and attains value that is very near to observational
bound. Again $\eta_{B}/s$ is plotted against the model parameters
$\phi_{0}$ and $V_{0}$ for different values of EOS parameter
$\omega$.  In both the cases, we have found results compatible
with observational data only for $\omega=1$. We get un-physical
results for the other values of $\omega$. We have compared the
results for $T_{D}=2\times10^{11}$ and for $T_{D}=2\times10^{16}$
and found that as $T_{D}$ gets lower values, $\eta_{B}/s$ gets
closer to the observational bound.

Next we have investigated the effect of a unified dark fluid model
GCG on the process of baryogenesis in the framework of
Ho$\check{r}$ava-Lifshitz gravity. We have observed that the
resulting baryon-to entropy significantly depends on the GCG model
parameters. It is observed that for $0.8\leq\alpha\leq1$ and
$T_{D}=2\times10^{11}$ the ratio gets values that are very close
to the observational bound. $\eta_{B}/s$ always decreases as
$\omega$ increases. In this scenario for $\omega<-1$,
$-1<\omega<0$ and $\omega=0$, the values of $\eta_{B}/s$ are in
very good agreement with observation, but $\omega<-1$ shows
un-physical result. Finally it must be stated that the quantum
mechanical nature of the background geometry of
Ho$\check{r}$ava-Lifshitz gravity plays an active role in the
observations obtained in this study.

\section*{Acknowledgements}

PR acknowledges University Grants Commission (UGC), Government of
India for providing research project grant (No. F.PSW-061/15-16
(ERO)). PR also acknowledges Inter University Centre for Astronomy
and Astrophysics (IUCAA), Pune, India, for awarding Visiting
Associateship.

\end{document}